\documentclass[a4paper, twoside, notitlepage,11pt]{article}
\usepackage[utf8]{inputenc}
\usepackage{natbib}
\usepackage{amsmath}
 \numberwithin{equation}{section}
\usepackage{float}        
\usepackage{hyperref}
\usepackage{amsfonts}
\usepackage{graphicx}
\usepackage{amssymb}
\usepackage{hyperref}
\usepackage{amsfonts}
\usepackage{graphicx}
\usepackage{caption}
\usepackage{array}
\usepackage[titletoc,toc,title]{appendix}
\usepackage{listings}
\usepackage{color}
\usepackage{titlesec}
\usepackage[titletoc,toc,title]{appendix}
\usepackage{verbatim}
\usepackage{setspace}
\usepackage{pdfpages}
\usepackage[margin=1.2in,footskip=1in]{geometry}

\titleformat{\section}{\bfseries}{\thesection.}{0.5em}{}
\titleformat{\subsection}{\normalfont\itshape}{\thesubsection.}{0.5em}{}
\titleformat{\subsubsection}{\normalfont\itshape}{\thesubsubsection.}{0.6em}{}

\renewcommand*\thesection{\arabic{section}} %slik at den starter kapittelnummerering på 1, ikke 0.1

\title{On the application of higher order symplectic integrators in Hamiltonian Monte Carlo}
\author{Janne Mannseth, Tore S. Kleppe, Hans J. Skaug}
\begin{document}
\pagenumbering{arabic}
\maketitle
    
\begin{abstract}
%Hamiltonian Monte Carlo (HMC) is an MCMC algorithm used to generate samples from complex probability distributions. 
We explore the construction of new symplectic numerical integration schemes to be used in Hamiltonian Monte Carlo and study their efficiency. 
Two integration schemes from \cite{bla2013numint}, and a new scheme based on optimal acceptance 
probability, are considered as candidates to the commonly used leapfrog method. 
All integration schemes are tested within the framework of the No-U-Turn sampler (NUTS),
both for a logistic regression model and a student $t$-model.
The results show that the leapfrog method is inferior to all the new methods both in terms of 
asymptotic expected acceptance probability for a model problem and the
and efficient sample size per computing time for the realistic models.
\end{abstract}


\section{Introduction}
Hamiltonian Monte Carlo \citep{due1987hybr, neal2010mcmc} (HMC) is a Markov chain Monte Carlo (MCMC) algorithm that
avoids the random walk behavior of the commonly used Metropolis-Hastings algorithm.
For this reason HMC has become popular for sampling from the posterior distribution
of Bayesian models \citep{liu2008montec,hof2011noutur}.
%p190 liu
%p 2 hoffman

HMC is applicable when 
the parameters have continuous distributions,
and requires that the gradient of the log-posterior can be evaluated.
Additional practical challenges with the HMC algorithm is its sensitivity to two user specified parameters;  
the so-called step size and the number of steps.
The NUTS algorithm \citep{hof2011noutur} automates the choice of these tuning parameters.
There exist extensions of HMC, such as Riemann Manifold HMC and Lagrangian HMC (see e.g. \cite{gir2011rieman}; \cite{shi2015mark}). 
Here we employ plain HMC because of its widespread use e.g. via the software package STAN \citep{car2015stan}.

An essential component of HMC is to evolve the Hamiltonian system derived from the posterior distribution 
while sampling is taking place. 
In practice the pair of differential equations that describes this system must be discretized, while retaining the
the volume-preserving and time-reverible nature of the exact solution \citep{neal2010mcmc}.
The most commonly used parametrized integration scheme used for HMC is the leapfrog method.
\cite{bla2013numint} have developed a framework for deriving more efficient integration schemes.
Such schemes have the potential to perform better than the leapfrog method, particularly in high dimensions. 
The leapfrog method often requires a small step size to obtian reasonable acceptance probabilites, whereas
the new integration schemes have the potentential to allow larger step sizes while retianing good acceptance
probabilites \citep{bla2013numint}. 

%We make use of a splitting method which ensures symplecticness and time-reversibility 
%of the numerical integrators. 
%More details can be found in \cite{bla2013numint}.

%\textcolor{blue}{Til Tore}: De to blå delene er vel akkurat det samme innholdet. Hvordan ville du ha løst dette?
%\textcolor{blue}{In order to measure the relative efficiency of the integration schemes, the trade-off between serial correlation in the sampled chain 
%and computation time is essential. 
%The efficient sample size (ESS) is a performance measure that can be used to this end.}
%Numerical integration schemes with more partial steps, or of a higher order, than the leapfrog method 
%can be expected to perform better with regards to acceptance probability. However, such schemes also require more of the costly 
%gradient evaluations.
%It is known that autocorrelation between states can be a problem when evaluating acceptance probabilities using the
%Metropolis-Hastings algorithm \citep{rob2001opti}. 
%For the Gaussian case the present paper uses a method that gives a proposal position vector that is independent of the previous 
%position vectors. 
%\textcolor{blue}{With low serial correlation it is generally so that a high acceptance probability is favorable, but not at the 
%expense of efficiency. Because of performance and the cost associated with evaluating the gradient 
%both of the measures acceptance probability and ESS are used to consider how efficient the various integrators are.}
The purpose of this paper is to compare the numerical integrators of \cite{bla2013numint}, and also
a new numerical integrator, to the leapfrog method
by considering realistic problems and experiments.
The effective sample size per computing time is used as the performance measure.
The results indicate that the new integrators perform better than the leapfrog method, with neglible
additional implementation effort.

The paper is laid out as follows: Section \ref{hmc} defines HMC and in Section~\ref{ints} the leapfrog method 
and the three other parameterized integration schemes are introduced.
An asymptotic expression for the expected acceptance probability is found for each of the 
numerical schemes based on a Gaussian test problem. 
Section~\ref{nutssec} embeds the 
integration schemes within the NUTS algorithm, and measures performance on a logistic
regression model and a student t-model. 
Lastly, Section~\ref{discuss} contains some discussion.
 
\section{Hamiltonian Monte Carlo}\label{hmc}
Denote the target (posterior) density by $\pi({q})$, where ${q} \in  \mathbb{R}^d$.
The system is augmented with another random vector ${p} \in \mathbb{R}^d$.
In the language of Hamiltonian dynamics $q(t)$ and $p(t)$ represent respectively the position and  momentum
of a particle at time $t$.
The Hamiltonian is denoted $H({q},{p})$ and can be expressed as \citep{neal2010mcmc}:
 \begin{equation} \label{hamiltonian}
  H({q},{p}) = U({q}) + \frac{1}{2}{p}^T{p},
\end{equation}
where $U({q}) = -\log(\pi({q}))$. The last part of equation (\ref{hamiltonian}) represents the negative log density of 
a $\mathcal{N}(0,I)$ distribution where the density is defined as $f(\textbf{x}) = (\frac{1}{\sqrt{2\pi}})^{\frac{n}{2}}
\exp(-\frac{1}{2}x^Tx)$ and $I$ is the identity matrix.
Note that strictly speaking only a kernel density proportional to the density is needed, not the density itself.
The Hamiltonian governs the evaluation of ${p(t)}$ and ${q(t)}$ via the following 
system of differential equations:
%\begin{align}\label{hamildiff}
%\begin{split}
 % \frac{d{q}}{dt} &= \frac{\partial H}{\partial {p}} = p, \\
 % \frac{d {p}}{dt} &= -\frac{\partial H}{\partial {q}} = - \nabla U(q) . 
%\end{split}
%\end{align}
\begin{equation}\label{hamildiff}
 \frac{d{q}}{dt} = \frac{\partial H}{\partial {p}} = p, \text{ } \text{ } \frac{d {p}}{dt} 
 = -\frac{\partial H}{\partial {q}} = - \nabla U(q),
\end{equation}
where $\nabla U(q)$ is the gradient of $ U(q)$. 

Let $(q,p)$ denote the current state of the sampler and $({q}^*, {p}^*)$ a proposal for the next. 
One iteration of the HMC algorithm involves the following steps. We start by drawing a new $p \sim \mathcal{N}(0,I)$. 
Next, we solve equation (\ref{hamildiff}) using $L$ steps of the integration scheme (leapfrog or 
a similar time-reversible, volume-preserving method). After $L$ steps
we arrive at the proposed position and  momentum, $q^*$ and $p^*$. 
The proposal is accepted \citep{due1987hybr, neal2010mcmc} with probability:
\begin{equation} \label{acceptanceprobability1}
  \alpha = \min(1,\exp(\Delta)),
  %=\text{min}(1, \text{exp} (-H({q}^*,{p}^*) + H({q},{p}))).
\end{equation}
where $\Delta=(-H({q}^*,{p}^*) + H({q},{p}))$. At each iteration $p^*$ is discarded after $\alpha$ has been computed.

\section{Symplectic numerical integration schemes}\label{ints}
As described above, volume-preserving and time-reversible integrators are essential to HMC.
Time is discretized in the differential equations (\ref{hamildiff}) by introducing the step size $\epsilon$.
Each parameterized numerical scheme consists of partial steps updating either position or momentum. For each step 
it is indicated where we are in the process of going from time $t$ to $t+\epsilon$. 
%Here we express the numerical integration schemes for time $t=0$.
%The steps in the integrator schemes are done for each $q_i$ and $p_i$, where $i=1,\dots,d$.

\subsection{The leapfrog method}
Consider the first integration scheme known as the leapfrog method \citep{lei2004simu, neal2010mcmc}, expressed for notational simplicity
for $t=0$:
%\begin{align}\label{leapfrogscheme}
% p(\epsilon/2) &= p(0) - \frac{\epsilon}{2}\frac{\partial U}{\partial q}(q(0)), \nonumber \\ 
% q(\epsilon) &= q(0) + \epsilon p(\epsilon/2),\\
% p(\epsilon) &= p(\epsilon/2)- \frac{\epsilon}{2}\frac{\partial U}{\partial q}(q(\epsilon)) \nonumber .
%\end{align}
\begin{align}\label{leapfrogscheme}
 p(\epsilon/2) &= p(0) - \frac{\epsilon}{2}\nabla U(q(0)), \nonumber \\ 
 q(\epsilon) &= q(0) + \epsilon p(\epsilon/2),\\
 p(\epsilon) &= p(\epsilon/2)- \frac{\epsilon}{2}\nabla U(q(\epsilon)) \nonumber .
\end{align}
Each partial step represents an update in either $p$ or $q$, 
jointly resulting in
an update for $(q,p)$ at time $\epsilon$. 
%If the proposal is accepted, $(q(\epsilon), p(\epsilon)) = (q^*, p^*)$, whereas $q(\epsilon) = q$ otherwise.
To obtain a proposal, (\ref{leapfrogscheme}) is repeated $L$ times, taking the output of the previous iteration as input to the next.
The leapfrog method is widely used in practice, and therefore 
constitutes a reasonable reference.

\subsection{Two-stage methods}
Following the method suggested in \cite{bla2013numint} the next parametrized integration scheme, composed of two leapfrog steps, is calculated to be as follows:
%\begin{align}\label{twostagescheme}
% q(a_1\epsilon) &= q(0) + a_1\epsilon p(0),\nonumber \\ 
% p(\epsilon/2) &= p(0) - \frac{\epsilon}{2}\frac{\partial U}{\partial q}(q(a_1\epsilon)),\nonumber \\
% q((1-a_1)\epsilon) &= q(a_1 \epsilon) - (1-2a_1)\epsilon p(\epsilon/2),\\
% p(\epsilon) &= p(\epsilon/2) - \frac{\epsilon}{2}\frac{\partial U}{\partial q}(q((1-a_1)\epsilon)),\nonumber \\
% q(\epsilon) &= q((1-a_1)\epsilon) + a_1\epsilon p(\epsilon). \nonumber 
%\end{align}
\begin{align}\label{twostagescheme}
 q(a_1\epsilon) &= q(0) + a_1\epsilon p(0),\nonumber \\ 
 p(\epsilon/2) &= p(0) - \frac{\epsilon}{2}\nabla U(q(a_1\epsilon)),\nonumber \\
 q((1-a_1)\epsilon) &= q(a_1 \epsilon) - (1-2a_1)\epsilon p(\epsilon/2),\\
 p(\epsilon) &= p(\epsilon/2) - \frac{\epsilon}{2}\nabla U(q((1-a_1)\epsilon)),\nonumber \\
 q(\epsilon) &= q((1-a_1)\epsilon) + a_1\epsilon p(\epsilon). \nonumber 
\end{align}
Following \cite{bla2013numint}, the value of the parameter is $a_1=\frac{3-\sqrt{3}}{6}$ (independent of the dimension $d$).
This value is found by minimizing an expression for the energy error $H(q,p)-H(q^*,p^*)$ with respect to $a_1$. 
Inserting this $a_1$ before performing any further calculations gives the integration scheme called the ``two-stage'' method.  

The next scheme has the same basis (\ref{twostagescheme}).
We suggest 
an alternative criterion for determining the
value for the parameter $a_1$.
%where its value is not immediately inserted into the scheme in (\ref{twostagescheme}). 
Now, $a_1$ is chosen so that the value of the expected acceptance probability $\alpha$ 
in (\ref{acceptanceprobability1}) is maximized. 
As shown in Appendix \ref{Appendix A} the optimal value is $a_1=\frac{3-\sqrt{5}}{4}$,
slightly smaller than for the two-stage method. Inserting this $a_1$ into equation (\ref{twostagescheme})
we get a new numerical scheme and name it the ``new two-stage'' method. The proposed method for finding the new value for $a_1$ is under the 
constraint that proposals are independent of the current state under a
standard $d$-dimensional Gaussian model. 

\subsection{The three-stage method}
The final integration scheme is called the three-stage method:
%\begin{align}\label{threestage}
% q(a_1\epsilon) &= q(0) + a_1\epsilon p(0), \nonumber \\ 
% p(b_1\epsilon) &= p(0) - b_1\epsilon \frac{\partial U}{\partial q}(q(a_1\epsilon)),\nonumber \\
% q(\epsilon/2) &= q(a_1\epsilon) + \left(\frac{1}{2}-a_1\right)\epsilon p(b_1\epsilon),\nonumber \\
% p((1-b_1)\epsilon) &= p(b_1\epsilon) - (1-2b_1)\epsilon\ \frac{\partial U}{\partial q}(q(\epsilon/2)),\\
% q((1-a_1)\epsilon) &= q(\epsilon/2) + \left(\frac{1}{2}-a_1\right)\epsilon p((1-b_1)\epsilon),\nonumber  \\
%p(\epsilon) &= p((1-b_1)\epsilon) - b_1\epsilon \frac{\partial U}{\partial q}(q((1-a_1)\epsilon)),\nonumber \\
%q(\epsilon) &= q((1-a_1)\epsilon) + a_1\epsilon p(\epsilon), \nonumber 
%\end{align}
\begin{align}\label{threestage}
 q(a_1\epsilon) &= q(0) + a_1\epsilon p(0), \nonumber \\ 
 p(b_1\epsilon) &= p(0) - b_1\epsilon \nabla U(q(a_1\epsilon)),\nonumber \\
 q(\epsilon/2) &= q(a_1\epsilon) + \left(\frac{1}{2}-a_1\right)\epsilon p(b_1\epsilon),\nonumber \\
 p((1-b_1)\epsilon) &= p(b_1\epsilon) - (1-2b_1)\epsilon\ \nabla U(q(\epsilon/2)),\\
 q((1-a_1)\epsilon) &= q(\epsilon/2) + \left(\frac{1}{2}-a_1\right)\epsilon p((1-b_1)\epsilon),\nonumber  \\
 p(\epsilon) &= p((1-b_1)\epsilon) - b_1\epsilon \nabla U(q((1-a_1)\epsilon)),\nonumber \\
 q(\epsilon) &= q((1-a_1)\epsilon) + a_1\epsilon p(\epsilon), \nonumber 
\end{align}
where the values of parameters $a_1$ and $b_1$ are found in \cite{bla2013numint} to be 
$a_1=\frac{12127897}{102017882}$ and $b_1=\frac{4271554}{14421423}$ using the energy error criterion.
All the introduces integration schemes are easy to implement as alternatives to the leapfrog method.
\subsection{Asymptotic acceptance probabilities}
Based on an $d$-dimensional standard Gaussian test problem, we compute asymptotic (in dimension) expressions for the 
expected acceptance probability when the integration times, $\epsilon L$, correspond to proposals that are independent
of the current state.

\noindent
[t]Table 1 near here [/t]

\noindent
Note that $\text{E}(\alpha)$ is $1+O(\epsilon^4)$ for the new two-stage while it is $1+O(\epsilon^2)$ for the other three. 
Hence, the new two-stage method has the highest acceptance rate for sufficiently small $\epsilon$,
owing to how the
value of parameter $a_1$ was found.
%Instead of inserting a value before the calculations, an asymptotic expression was optimized 
%with respect to $a_1$. 
%Then further calculations was carried out using this new $a_1$, namely 
%\begin{equation} a_1=\frac{3-\sqrt{5}}{4}.\end{equation}
%This procedure gives an expression of a higher order. 

\cite{bes2010optim} provide a general result that shows that $\epsilon$ should behave as $O(d^{-\frac{1}{4}}).$ 
To obtain $\text{E}(\Delta)=O(1)$ for $L\epsilon=O(1)$, the results in Table 1 also indicate $\epsilon=O(d^{-\frac{1}{4}})$ for  
leapfrog, two-stage and three-stage under the standard Gaussian model problem.
If $\epsilon$ decreases faster we will waste computation, and if it decreases slower,
 the MCMC chain will stagnate and be ineffective.
On the other hand, the acceptance probability for the new two-stage method imply $\epsilon=O(d^{-\frac{1}{8}})$ 
to obtain $\text{E}(\Delta)=O(1)$ for $L\epsilon=O(1)$. 
This could indicate that the new two-stage 
numerical scheme is too specialized to the particular model problem used to find it, and will be furhter explored
in section 4. 
More details on the calculations in Table 1 are given in Appendix \ref{Appendix A}.

\subsection{Comparing performance on the model problem}
To further add understanding of the results provided in Table 1, we compare a 
%CPU-time 
weighted measure of 
efficiency based on $\text{E}(\alpha)$ for different numbers of dimensions. Again, we use a $d$-dimensional standard Gaussian 
model problem.
 Let $\Upsilon$ denote the expected number of accepted movements per calculation time, as a measure of efficiency. 
 For the leapfrog method 
 we have $\Upsilon = \frac{\text{E}(\alpha)}{L}$, for the
 two-stage methods $\Upsilon = \frac{\text{E}(\alpha)}{2L}$, where the factor $2$ in the denominator is
 an effect of increase in number of gradient evaluations. 
 Correspondingly
 the three-stage method has $\Upsilon = \frac{\text{E}(\alpha)}{3L}$. The measure is considered for different values of dimension $d$
 and the maximum $\Upsilon$ within each dimension is the preferable one. 
For each dimension the
 \begin{equation}\label{maksupsi}
  %\max_{\epsilon}\lim_{d\to\infty} \Upsilon(\epsilon,d)
   \max_{\epsilon} \Upsilon(\epsilon,d),
 \end{equation}
 where $\epsilon$ is such that $L\epsilon \approx \frac{\pi}{2}$ corresponds to independent proposals,
 is found for all four numerical schemes.
 For each of the numerical schemes, Figure 1 shows how the maximum $\Upsilon$ develops as $d$ increases. 
 We see that the leapfrog method 
 starts out the best, but is quickly overtaken by all the other schemes as $d$ increases. 
 It is also seen that the performance of the new two-stage method, 
 which is of a higher order than the other schemes, stands out for high $d$.
 Figure 2 shows the ratio between $\Upsilon$ for the leapfrog method 
 and each of the other numerical schemes. 
 For each fraction, a value greater than one indicates that 
 the suggested method (for the corresponding $d$) is the best of the two methods considered. In Figure 2 
 we see a different illustration on how the 
 new methods perform better than
 leapfrog method as $d$ increases. It is clear that the new two-stage method is the most efficient of all methods considered here .

[f]Figure 1 near here [/f]
% \begin{figure}
%  \centering
%  \includegraphics[width=13cm]{eff.jpeg}
%  \caption{Logarithm of maximum $\Upsilon$ as a function of log($d$)}
%  \label{fig:maxupsall}
%\end{figure}

[f]Figure 2 near here [/f]
% \begin{figure}
%  \centering
%  \includegraphics[width=13cm]{compeff.jpeg}
%  \caption{Relations between maximum $\Upsilon$ as a function of log($d$)}
%  \label{fig:compupsall}
%\end{figure}
% \noindent 

Based on these results the way forward is to determine whether this is relevant in practice. Thus we need to consider some 
 realistic examples. 
 
 \section{Simulation studies and realistic applications}\label{nutssec}
 Arguably, the $d$-dimensional standard Gaussian problem (which Table 1 relies on) is too simple for 
 us to claim the new integrators constitute improvements over leapfrog in practice. This section considers realistic models.
 In the $d$-dimensional Gaussian case, we used proposals independent of current state.
 In a general case that does not lend it self to analytic solution of (\ref{hamildiff}), we resort to the No-U-Turn sampler for choosing 
 appropriate integration times.
 
 \subsection{The No-U-Turn Sampler}
   The No-U-Turn Sampler (NUTS) was introduced by \cite{hof2011noutur} as an improvement to the HMC algorithm.
  One reason for using NUTS here is to avoid the 
   sensitivity to user specified parameters, because both too low and too high values of step size $\epsilon$ and number of steps $L$
   can cause problems \citep{hof2011noutur}. 
   The NUTS algorithm discovers the point where the HMC algorithm could start to explore random walk behaviour by allowing no u-turns in 
   the trajectory. This way the choice of $L$ is eliminated.
   In addition, a method for adapting the step size $\epsilon$ by using dual averaging 
   is implemented in NUTS \citep{hof2011noutur}. Using NUTS with adaptive selection of $\epsilon$ eliminates all user
   specified choices from Section \ref{ints}. In the case of the four numerical schemes, this means that their 
   efficiency can be tested without user specified parameters. All subsequent computations are carried out in MATLAB.
    
\subsection{Simulation testing using NUTS and logistic regression model}\label{logreg}
As the first test case, we use a Bayesian logistic regression model, and the 
collection of five data sets and a Bayesian logistic regression model
from \cite{gir2011rieman}. The five data sets vary in number of observations $n$ from $250$ to $1000$ and 
in number of parameters 
$d$ from $7$ to $25$. 
The model layout is identical to \cite{gir2011rieman} where the prior is given as 
$\boldsymbol\beta \sim \mathcal{N}(\textbf{0}, 100 \textbf{I})$.
The effective sample size (ESS) is found by following
\cite{gir2011rieman} and used in the performance measurement ESS/CPU time. 
All the experiments are repeated ten times, so that 
the results in Table 2 are mean results over these ten runs. Table
2 presents the mean CPU times, the mean minimum, median and 
maximum ESS and the mean number of step sizes for all 
numerical integration schemes and data sets. 
The final two columns of Table 2 show minimum ESS/CPU 
time and median ESS/CPU time. The number of samples were 5000 with burn-in 
set to 1000 for all experiments. 

\noindent
[t]Table 2 near here [/t]

\noindent
\noindent
In general Table 2 shows that the 
higher order numerical schemes get higher ESS/CPU time than the leapfrog method. 
The two-stage method generally performs better than the new two-stage method.
Four data sets have ESS that reach the number of samples (5000). 
This means that there is little or no autocorrelation between the samples.
Table 2 shows that the three-stage method or the two-stage method perform the best
within each data set, as they have the largest minimum 
ESS/CPU time. The leapfrog 
method has the lowest value of minimum ESS/CPU time for all five data sets. 
The last column of Table 2, the median ESS/CPU time, confirms the same pattern 
as for the minimum ESS/CPU time.
It is expected that the step size increases in relation to the number of steps the methods have.
With regards to step size we have that 
$\text{the two-stage method}$ $>$ (the leapfrog method)/2 for all data sets except "Ripley``.
The same goes for the new two-stage method. Correspondingly we 
have that the three-stage method $>$ (the leapfrog method)/3 with regards to step size. For the ``Ripley'' 
data set this only holds for the new two-stage method, while the other two are very close.
With these results we see that the suggested integration schemes outperform the leapfrog method also when considering a more realistic problem 
than the standard Gaussian test problem. 

\subsection{Simulation testing using NUTS and student t-model}
In addition to the logistic model in Section \ref{logreg}, we also consider a multivariate student t distribution. 
The target density kernel is given as
\[\pi(x) \propto \left(1 + \frac{1}{\nu}x^T \Sigma^{-1}x\right)^{-\frac{\nu+d}{2}},\]
where the degrees of freedom, $\nu$, is set to $5$ and dimension $d$ is $2,10$ and $100$. 
$\Sigma^{-1}$ is the precision matrix associated with a Gaussian AR($1$) model with 
autocorrelation $0.95$ and unit innovation variance.
All variations are run ten times and the results in
Table 3 are the means taken over these ten runs. 

\noindent
[t]Table 3 near here [/t]

\noindent
\begin{comment}
\begin{table}[H]
\centering
\resizebox{1.07 \textwidth}{!}{\begin{tabular}{l c c c c c c c}
Numerical scheme &  \shortstack{CPU time\\ (s)} &  \shortstack{Integrator\\ step size} &\shortstack{minimum \\ESS}& \shortstack{median\\ ESS} &\shortstack{maximum \\ESS}
&\shortstack{min ESS / \\ CPU time} &\shortstack{med ESS / \\ CPU time} \\
\hline
Dimension $d=2$ \\
Leapfrog & 13.80 & 0.7897 & 859 & 876 & 893 & \textbf{62.25} & \textbf{63.48} \\
Two-stage & 18.34 & 1.5863 & 962 & 969 & 977 & 52.45 & 52.84 \\
New two-stage & 20.41 & 1.3941 & 963 & 975 & 987 & 47.18 & 47.77 \\
Three-stage & 15.90 & 2.3257 & 955 & 963 & 971 & 60.06 & 60.57 \\ 

Dimension $d=10$ \\
Leapfrog & 48.59 & 0.4027 & 577 & 633 & 747 & 11.87 & 13.03 \\
Two-stage & 59.89  & 0.8027 & 605 & 645 & 748 & 10.10 & 10.77 \\
New two-stage & 69.54 & 0.7027 & 614 & 651 & 749 & 8.83 & 9.36 \\
Three-stage & 54.68 & 1.2399 & 696 & 731 & 845 & \textbf{12.72} & \textbf{13.37} \\

Dimension $d=100$ \\
Leapfrog & 853 & 0.2615 & 1180 & 1552 & 2416 & 1.38 & 1.82 \\
Two-stage & 1021 & 0.6171 & 1156 & 1533 & 2372 & 1.13 & 1.50 \\
New two-stage & 1162 & 0.5647 & 1242 & 1537 & 2353 & 1.07 & 1.32 \\
Three-stage & 699 & 1.0762 & 1228 & 1521 & 2333 & \textbf{1.75} & \textbf{2.18} \\
\end{tabular}}
\caption{ESS and CPU times when using the four numerical schemes with a $d$-dimensional student t-model. The values in bold face 
represent the best result within each dimension.}
\label{table:studtt_table}
\end{table} 
\end{comment}

\noindent
From Table 3 we see that in general the 
two-stage method performs better than the new two-stage method.
For each dimension the highest minimum and median ESS is obtained with  
the two-stage method or the three-stage method. 
However, because of the time cost for the two-stage methods, the leapfrog method gets 
a slightly higher value of $\frac{\text{min ESS}}{\text{CPU time}}$ and $\frac{\text{med ESS}}{\text{CPU time}}$. 
The three-stage method requires less 
CPU time and gives the highest $\frac{\text{min ESS}}{\text{CPU time}}$ and $\frac{\text{med ESS}}{\text{CPU time}}$ 
of all four numerical integrators for dimensions $d=10$ and $d=100$. 
We see that for $d=100$
that the step sizes of the two-stage methods and the three-stage method are greater than the step size of $\frac{\text{the leapfrog method}}{2}$ and 
 $\frac{\text{the leapfrog method}}{3}$ respectively. For $d=10$ this only holds for the three-stage method. For $d=2$ it holds for the 
 two-stage method.
 Considering that the integration schemes are all easy to implement and work with, 
 these results give reason to further explore the use of them.

\section{Discussion}\label{discuss}
This paper makes the contribution of introducing and developing numerical schemes as alternatives to the leapfrog method. For a $d$-dimensional standard 
Gaussian model all three new numerical integration schemes perform better than the original scheme as dimension $d$ increases. It is seen 
that a numerical integration scheme of a higher order, like 
the new two-stage method, stands out for the Gaussian model, especially for larger~$d$. 
Using these schemes can require more costly computations because of the number of gradient evaluations. 
However, they are still superior to their precursor in HMC. 

For the logistic regression model 
%(\citep{gir2011rieman}) 
the new 
schemes were all shown to be more efficient than the leapfrog method. Even with the increased gradient evaluations the 
results were positive 
regarding the efficiency of the suggested numerical integration schemes, for all the five datasets considered. 
%We see that introducing alternatives to the leapfrog method in HMC is appropriate.

For the student t-model the suggested numerical integration schemes all obtain higher minimum and median ESSes than the leapfrog 
method for $d=2$ and $d=10$. For $d=100$, the ESSes are quite close, but only the three-stage method has a higher minimum ESS than the leapfrog method. 
The leapfrog method has a higher minimum and median ESS/CPU time than the two-stage methods.
This can be because of the cost related to the gradient evaluations of the two-stage methods, which can impact the 
CPU time. Still, the tree-stage method is the most efficient of all schemes as $d$ increases. This means that also for the student t-model there 
can be efficiency gain by replacing the leapfrog method with other numerical schemes.

These results give reason to explore the development of numerical schemes on the form seen in this paper. 
By doing this the performance and efficiency of 
HMC can be improved, especially as the dimension increases. 

\newpage

%%%%%%%%%%%%%%%%%%%%%%%%%%%%%%%%%%%%%% This is where the appendix starts %%%%%%%%%%%%%%%%%%%%%%%%%%%
\titleformat{\section}{\large\bfseries}{\appendixname~\thesection .}{0.5em}{}
\begin{appendices}
\section{Details}\label{Appendix A}
The $U(q)$ used when obtaining the results in Section~\ref{ints} is
$U(q) = \frac{1}{2}q^Tq$ and the Hamiltonian then is:
\[H = \frac{1}{2}q^Tq + \frac{1}{2}p^Tp,\]
where $q,p \in \mathbb{R}^d$.
Owing to the fact that the gradient of $H$ with respect to both $q$ and $p$ are linear, $H$ 
together with a numerical integration scheme gives a $2 \times 2$ matrix $M_{\epsilon}$ (one for each scheme) 
whose elements change with 
step size $\epsilon$ so that 
\begin{equation}\label{matleap}
  \begin{bmatrix}
    q_i(\epsilon) \\
    p_i(\epsilon)
  \end{bmatrix}
  =M_{\epsilon}
  \begin{bmatrix}
    q_i(0) \\
    p_i(0)
  \end{bmatrix}, \;i=1,\dots,d.
 \end{equation}
 Using the leapfrog scheme as an example, the matrix $M_\epsilon$ in (\ref{matleap}) is
\[ A= \begin{bmatrix}
      1-\epsilon^2/2 & \epsilon \\
      -\epsilon + \epsilon^3/4 & 1- \epsilon^2/2
     \end{bmatrix}.\]
The chosen integration scheme is applied $L$ times, so that the each element of the proposal obtains as
  \begin{equation*}\label{posLL}
  \begin{bmatrix}
    q_i^* \\
    p_i^*
  \end{bmatrix}
  =M_{\epsilon}^L\begin{bmatrix}
    q_i(0) \\
    p_i(0)
  \end{bmatrix},\;i=1,\dots,d.
 \end{equation*}
To obtain proposals that are independent of the current state, we consider combinations of 
$\epsilon$ and $L$ so that the diagonal elements of $M_{\epsilon}^L$ are zero (i.e. $\epsilon L\approx \pi/2$),
and get:
 \begin{equation*}
  \begin{bmatrix}
    q_i^* \\
    p_i^*
  \end{bmatrix}
  =
  \begin{bmatrix}
    0 & r_L \\
    s_L & 0
  \end{bmatrix}
  \begin{bmatrix}
    q_i(0) \\
    p_i(0)
  \end{bmatrix}
 \end{equation*}
For the leapfrog method, the $r_L$ and $s_L$ are:
\begin{align*}
 r_L &= 1 + \frac{1}{8}\epsilon^2 + \frac{3}{128}\epsilon^4 + \frac{5}{1024}\epsilon^6 + O(\epsilon^8),\\
s_L &= -1 + \frac{1}{8}\epsilon^2 + \frac{1}{128}\epsilon^4 + \frac{1}{1024}\epsilon^6 + O(\epsilon^8).
\end{align*}
Now, recall that the acceptance probability $\alpha$:
\begin{equation*} \label{acceptanceprobability}
  \alpha = \text{min}(1, \text{exp} (\Delta)), 
\end{equation*}
where
$\Delta =(-H(q^*,p^*) + H(q,p)).$
%Recall that \[\alpha=\text{min}(1,\text{exp}\Delta)=\text{exp}(\text{min}(0,\Delta)).\]
For dimension $d$,  $\Delta$ simplifies as:
\begin{equation*}
\Delta  
= \frac{1}{2}[\sum_{i=1}^{d}[(1-s_L^2)q_i^2 + (1-r_L^2)p_i^2]].
\end{equation*}
Note that from expressions for $r_L$ and $s_L$, we have that $r_L \rightarrow 1$ and $s_L \rightarrow -1$ 
when $L \rightarrow \infty$. As a result of this, $\Delta \rightarrow 0$. 
This behaviour is in correspondence with what we expect, because it indicates that the acceptance probability will go towards one.

Owing to the fact that $q_i,\;p_i,\;i=1,\dots,d$ are iid standard normal,
expressions for $\mu = \text{E}(\Delta)$ and $\sigma^2 = \text{Var}(\Delta)$ can be found for each
of the integration schemes. 
Further, we use the central limit theorem so that 
\[\frac{\Delta - \mu}{\sigma} \xrightarrow[d\rightarrow \infty]{} N(0,1),\]
and this forms the basis for the formulas in Table 1.

\end{appendices}
%\addtocontents{toc}{\protect\contentsline{chapter}{Appendix:}{}}
 
 \clearpage
\bibliographystyle{chicago}
\bibliography{bibl}

\begin{thebibliography}{}

\bibitem[\protect\citeauthoryear{Beskos, Pillai, Roberts, Sanz-Serna, Stuart,
  et~al.}{Beskos et~al.}{2013}]{bes2010optim}
Beskos, A., N.~Pillai, G.~Roberts, J.-M. Sanz-Serna, A.~Stuart, et~al. (2013).
\newblock Optimal tuning of the hybrid monte carlo algorithm.
\newblock {\em Bernoulli\/}~{\em 19\/}(5A), 1501--1534.

\bibitem[\protect\citeauthoryear{Blanes, Casas, and Sanz-Serna}{Blanes
  et~al.}{2014}]{bla2013numint}
Blanes, S., F.~Casas, and J.~Sanz-Serna (2014).
\newblock {Numerical integrators for the hybrid Monte Carlo method}.
\newblock {\em SIAM Journal on Scientific Computing\/}~{\em 36\/}(4),
  A1556--A1580.

\bibitem[\protect\citeauthoryear{Carpenter, Gelman, Hoffman, Lee, Goodrich,
  Betancourt, Brubaker, Guo, Li, and Riddell}{Carpenter
  et~al.}{2015}]{car2015stan}
Carpenter, B., A.~Gelman, M.~Hoffman, D.~Lee, B.~Goodrich, M.~Betancourt, M.~A.
  Brubaker, J.~Guo, P.~Li, and A.~Riddell (2015).
\newblock Stan: a probabilistic programming language.
\newblock {\em Journal of Statistical Software\/}.

\bibitem[\protect\citeauthoryear{Duane, Kennedy, Pendleton, and Roweth}{Duane
  et~al.}{1987}]{due1987hybr}
Duane, S., A.~D. Kennedy, B.~J. Pendleton, and D.~Roweth (1987).
\newblock {Hybrid Monte Carlo}.
\newblock {\em Physics letters B\/}~{\em 195\/}(2), 216--222.

\bibitem[\protect\citeauthoryear{Girolami and Calderhead}{Girolami and
  Calderhead}{2011}]{gir2011rieman}
Girolami, M. and B.~Calderhead (2011).
\newblock Riemann manifold langevin and hamiltonian monte carlo methods.
\newblock {\em Journal of the Royal Statistical Society: Series B (Statistical
  Methodology)\/}~{\em 73\/}(2), 123--214.

\bibitem[\protect\citeauthoryear{Hoffman and Gelman}{Hoffman and
  Gelman}{2014}]{hof2011noutur}
Hoffman, M.~D. and A.~Gelman (2014).
\newblock {The No-U-Turn Sampler: adaptively setting path lengths in
  Hamiltonian Monte Carlo.}
\newblock {\em Journal of Machine Learning Research\/}~{\em 15\/}(1),
  1593--1623.

\bibitem[\protect\citeauthoryear{Lan, Stathopoulos, Shahbaba, and Girolami}{Lan
  et~al.}{2015}]{shi2015mark}
Lan, S., V.~Stathopoulos, B.~Shahbaba, and M.~Girolami (2015).
\newblock Markov chain monte carlo from lagrangian dynamics.
\newblock {\em Journal of Computational and Graphical Statistics\/}~{\em
  24\/}(2), 357--378.

\bibitem[\protect\citeauthoryear{Leimkuhler and Reich}{Leimkuhler and
  Reich}{2004}]{lei2004simu}
Leimkuhler, B. and S.~Reich (2004).
\newblock {\em {Simulating Hamiltonian dynamics}}, Volume~14.
\newblock Cambridge University Press.

\bibitem[\protect\citeauthoryear{Liu}{Liu}{2008}]{liu2008montec}
Liu, J.~S. (2008).
\newblock {\em Monte Carlo strategies in scientific computing}.
\newblock Springer Science \& Business Media.

\bibitem[\protect\citeauthoryear{Neal}{Neal}{2010}]{neal2010mcmc}
Neal, R.~M. (2010).
\newblock {MCMC using Hamiltonian dynamics}.
\newblock {\em Handbook of Markov Chain Monte Carlo\/}~{\em 54}, 113--162.

\end{thebibliography}
\addcontentsline{toc}{chapter}{References}

\newpage
 \begin{figure}[H]
  \centering
  \includegraphics[width=13cm]{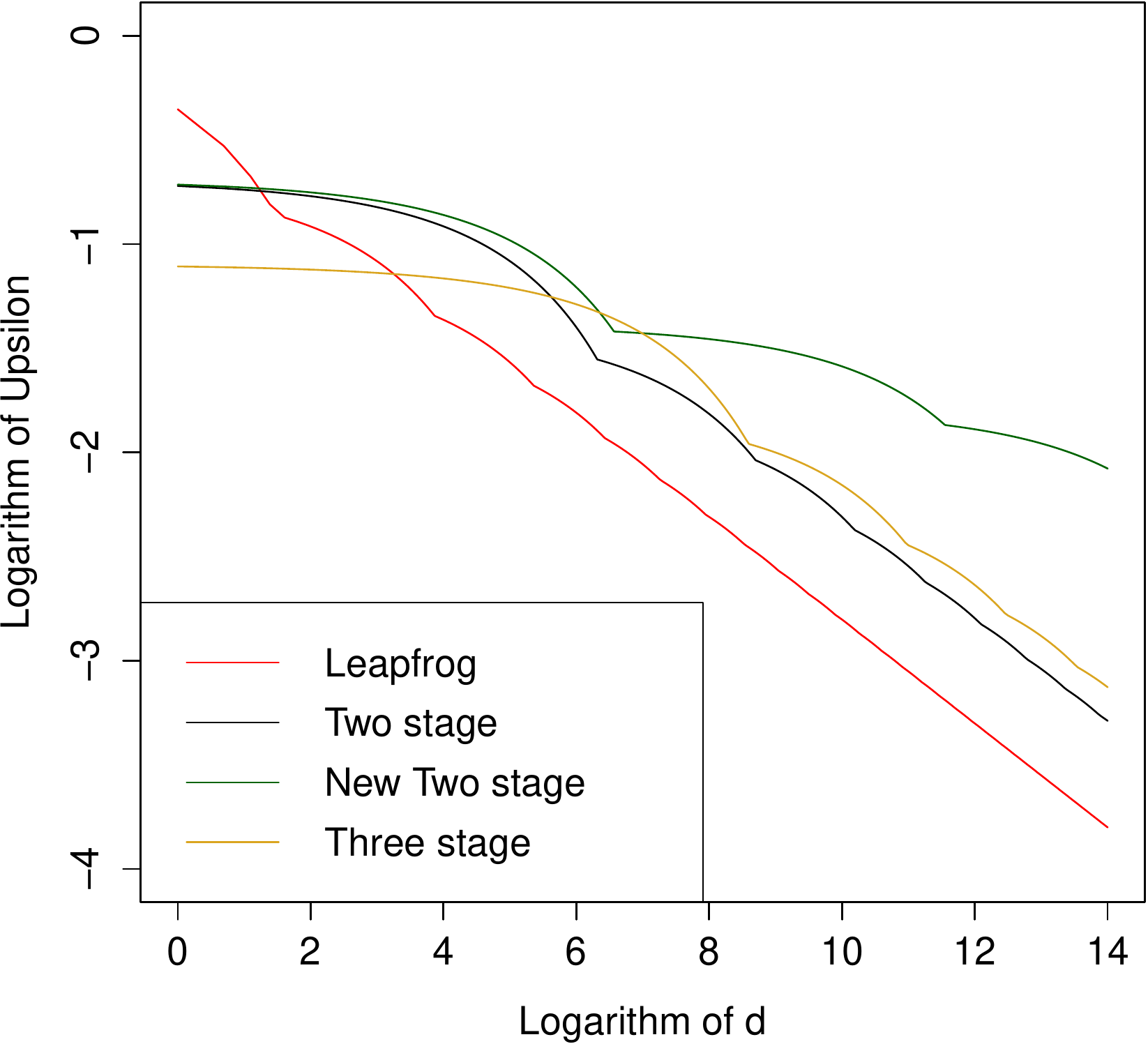}
  \caption{Logarithm of maximum expected number of accepted movements per calculation time as a function of log($d$).}
  \label{fig:maxupsall}
\end{figure}
\newpage
 \begin{figure}[H]
  \centering
  \includegraphics[width=13cm]{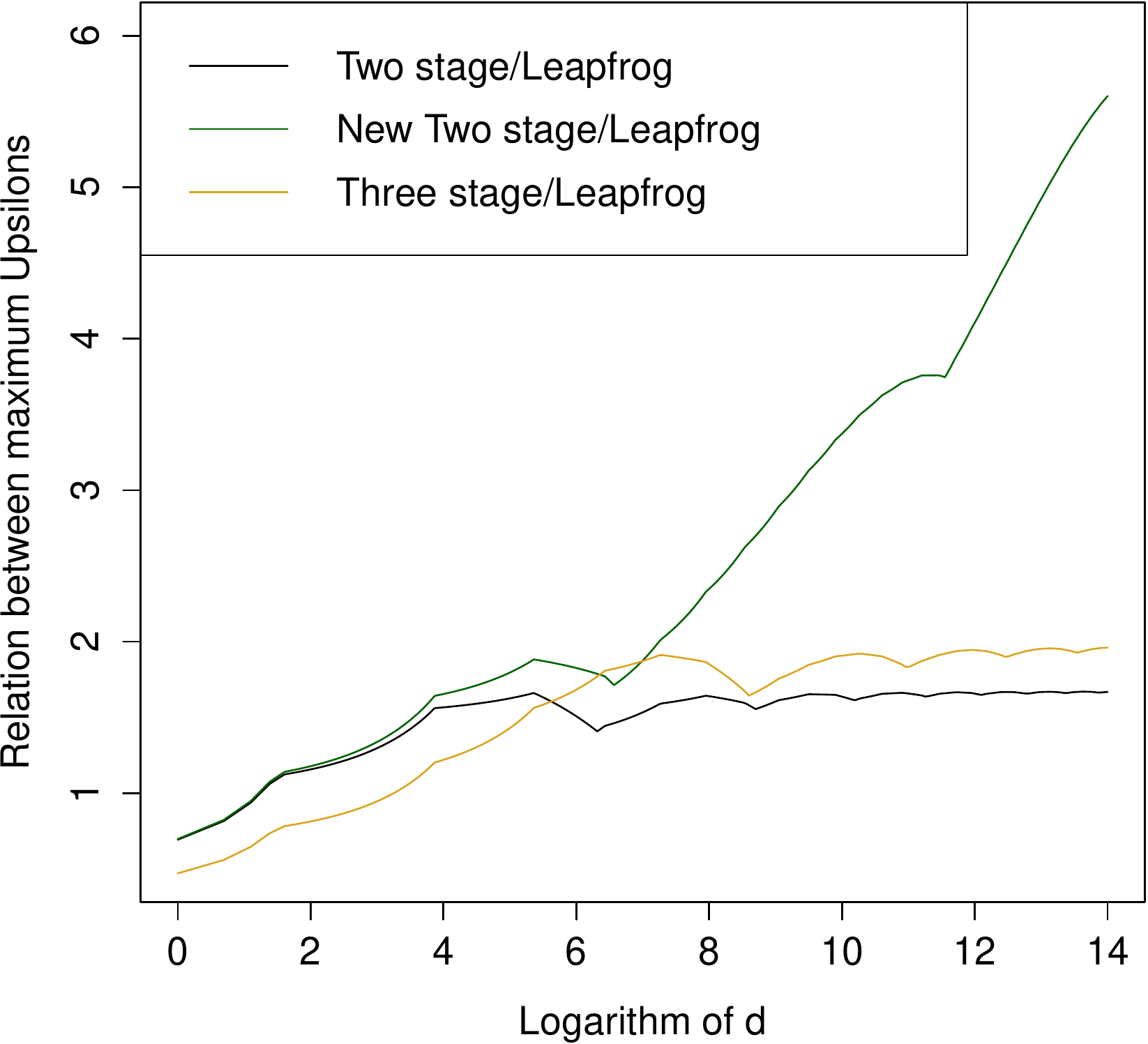}
  \caption{Relations between maximum expected number of accepted movements per calculation time as a function of log($d$).}
  \label{fig:compupsall}
\end{figure}
\newpage
\subsection*{Table 1:}
\includegraphics[width=13cm]{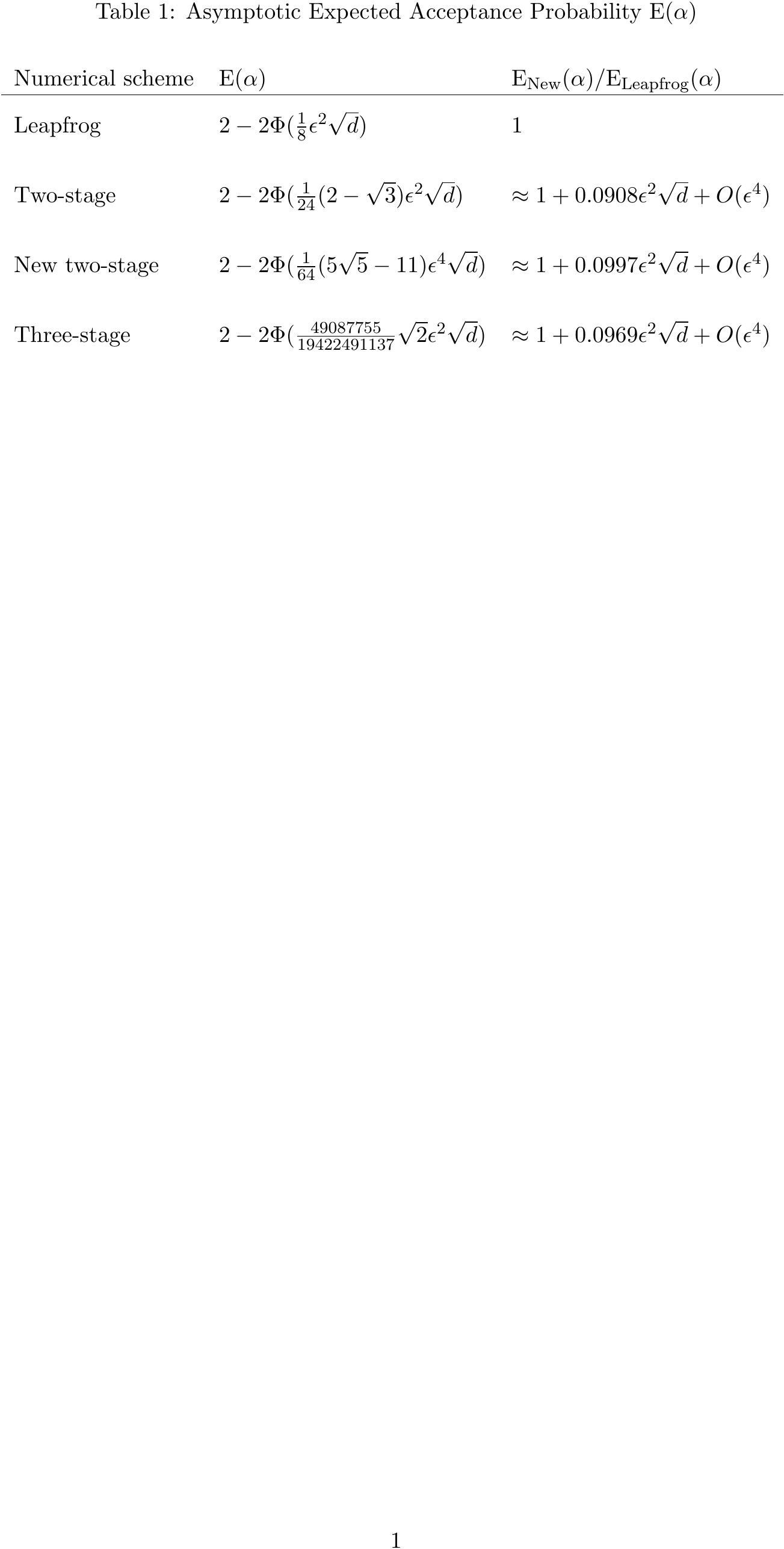}
\newpage
\subsection*{Table 2:}
\includegraphics[width=13cm]{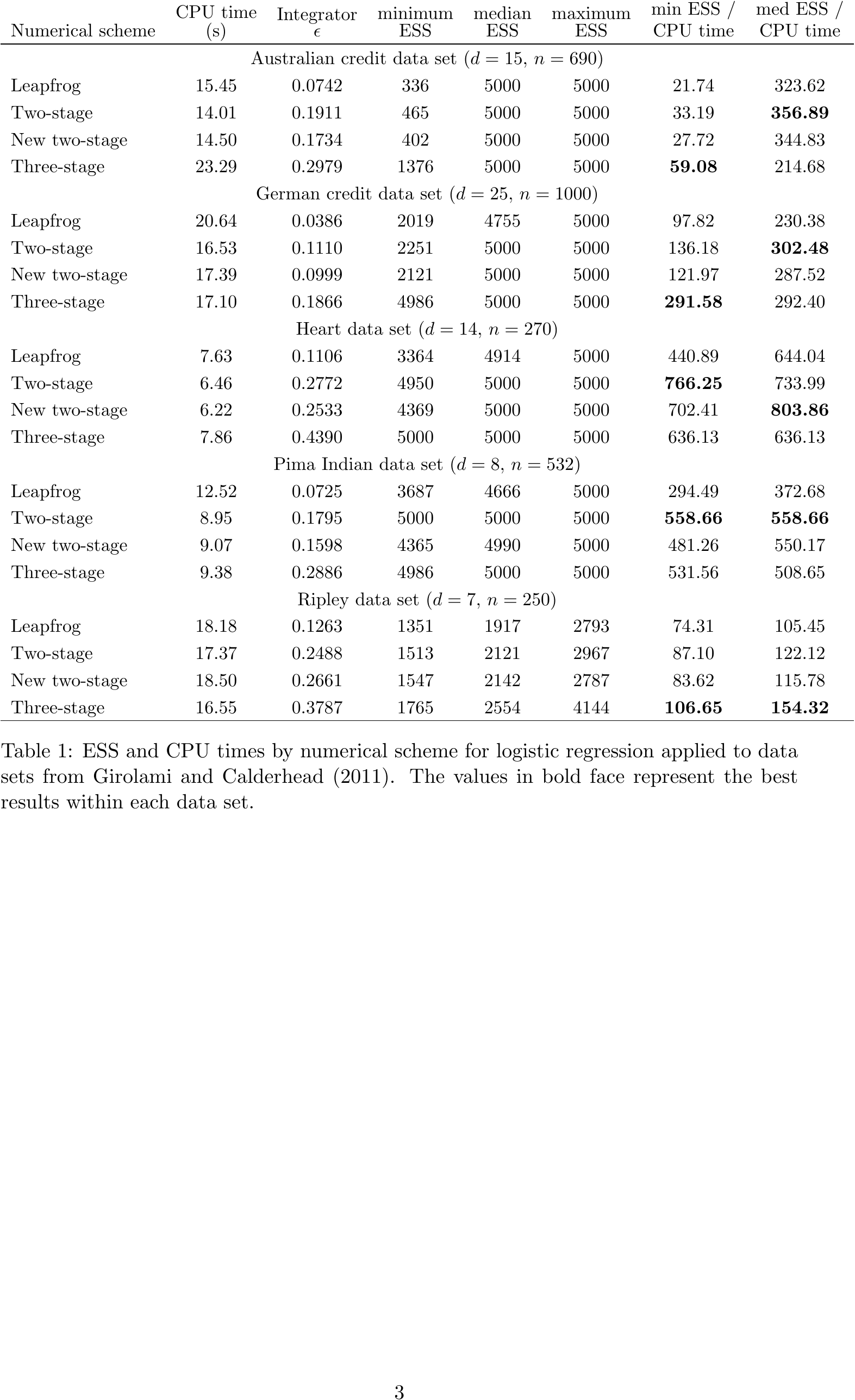}
\newpage
\subsection*{Table 3:}
\includegraphics[width=13cm]{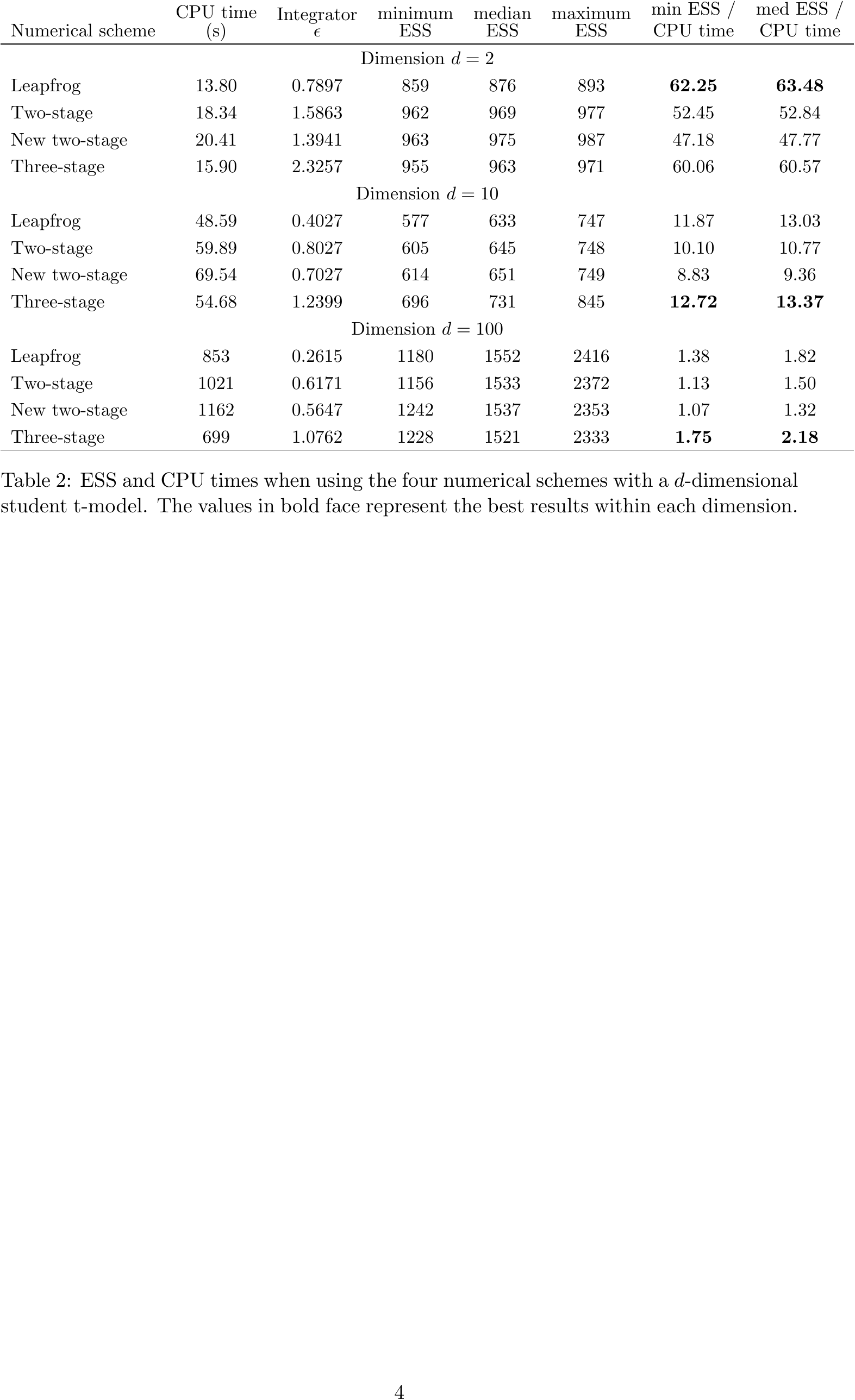}
\end{document}